\documentclass[10pt,twocolumn,twoside]{IEEEtran}

\usepackage{amssymb}
\usepackage{amsfonts}
\usepackage{amsmath}
\usepackage{cite}
\usepackage[dvips]{graphicx}
\usepackage{color}
\usepackage{comment}
\usepackage{makecell}

\begin{document}
\newtheorem{definition}{\it Definition}
\newtheorem{theorem}{\bf Theorem}
\newtheorem{lemma}{\it Lemma}
\newtheorem{corollary}{\it Corollary}
\newtheorem{remark}{\it Remark}
\newtheorem{example}{\it Example}
\newtheorem{case}{\bf Case Study}
\newtheorem{assumption}{\it Assumption}
\newtheorem{property}{\it Property}
\newtheorem{proposition}{\it Proposition}

\newcommand{\hP}[1]{{\boldsymbol h}_{{#1}{\bullet}}}
\newcommand{\hS}[1]{{\boldsymbol h}_{{\bullet}{#1}}}

\newcommand{\ba}{\boldsymbol{a}}
\newcommand{\baq}{\overline{q}}
\newcommand{\bA}{\boldsymbol{A}}
\newcommand{\bb}{\boldsymbol{b}}
\newcommand{\bB}{\boldsymbol{B}}
\newcommand{\bc}{\boldsymbol{c}}
\newcommand{\bcO}{\boldsymbol{\cal O}}
\newcommand{\bh}{\boldsymbol{h}}
\newcommand{\bH}{\boldsymbol{H}}
\newcommand{\bl}{\boldsymbol{l}}
\newcommand{\bm}{\boldsymbol{m}}
\newcommand{\bn}{\boldsymbol{n}}
\newcommand{\bo}{\boldsymbol{o}}
\newcommand{\bO}{\boldsymbol{O}}
\newcommand{\bp}{\boldsymbol{p}}
\newcommand{\bq}{\boldsymbol{q}}
\newcommand{\bR}{\boldsymbol{R}}
\newcommand{\bs}{\boldsymbol{s}}
\newcommand{\bS}{\boldsymbol{S}}
\newcommand{\bT}{\boldsymbol{T}}
\newcommand{\bw}{\boldsymbol{w}}

\newcommand{\balpha}{\boldsymbol{\alpha}}
\newcommand{\bbeta}{\boldsymbol{\beta}}
\newcommand{\bOmega}{\boldsymbol{\Omega}}
\newcommand{\bTheta}{\boldsymbol{\Theta}}
\newcommand{\bphi}{\boldsymbol{\phi}}
\newcommand{\btheta}{\boldsymbol{\theta}}
\newcommand{\bvarpi}{\boldsymbol{\varpi}}
\newcommand{\bpi}{\boldsymbol{\pi}}
\newcommand{\bpsi}{\boldsymbol{\psi}}
\newcommand{\bxi}{\boldsymbol{\xi}}
\newcommand{\bx}{\boldsymbol{x}}
\newcommand{\by}{\boldsymbol{y}}

\newcommand{\cA}{{\cal A}}
\newcommand{\bcA}{\boldsymbol{\cal A}}
\newcommand{\cB}{{\cal B}}
\newcommand{\cE}{{\cal E}}
\newcommand{\cG}{{\cal G}}
\newcommand{\cH}{{\cal H}}
\newcommand{\bcH}{\boldsymbol {\cal H}}
\newcommand{\cK}{{\cal K}}
\newcommand{\cO}{{\cal O}}
\newcommand{\cR}{{\cal R}}
\newcommand{\cS}{{\cal S}}
\newcommand{\dcS}{\ddot{{\cal S}}}
\newcommand{\ds}{\ddot{{s}}}
\newcommand{\cT}{{\cal T}}
\newcommand{\cU}{{\cal U}}
\newcommand{\wt}[1]{\widetilde{#1}}

\newcommand{\mA}{\mathbb{A}}
\newcommand{\mE}{\mathbb{E}}
\newcommand{\mG}{\mathbb{G}}
\newcommand{\mR}{\mathbb{R}}
\newcommand{\mS}{\mathbb{S}}
\newcommand{\mU}{\mathbb{U}}
\newcommand{\mV}{\mathbb{V}}
\newcommand{\mW}{\mathbb{W}}

\newcommand{\uq}{\underline{q}}
\newcommand{\ubq}{\underline{\boldsymbol q}}

\newcommand{\red}[1]{\textcolor[rgb]{1,0,0}{#1}}
\newcommand{\gre}[1]{\textcolor[rgb]{0,1,0}{#1}}
\newcommand{\blu}[1]{\textcolor[rgb]{0,0,0}{#1}}

\title{From Semantic Communication to Semantic-aware Networking: Model, Architecture, and Open Problems} 

\author{
Guangming~Shi, 
Yong~Xiao, 
Yingyu~Li, Xuemei~Xie 

%
\thanks{

G. Shi and X. Xie are with School of Artificial Intelligence, Xidian University, Xi'an, Shaanxi 710071, China (e-mail: \{gmshi, xmxie\}@xidian.edu.cn). G. Shi is also with Pazhou Lab, Guangzhou, China.

Y. Xiao and Y. Li are with the School of Electronic Information and Communications at the Huazhong University of Science and Technology, Wuhan, China 430074 (e-mail: \{yongxiao, liyingyu\}@hust.edu.cn). Y. Xiao is also with Pazhou Lab, Guangzhou, China.

}
}





\maketitle

\begin{abstract}
Existing communication systems are mainly built based on Shannon's information theory which deliberately ignores the semantic aspects of communication. The recent iteration of wireless technology, the so-called 5G and beyond, promises to support a plethora of services enabled by carefully tailored network capabilities based on contents, requirements, as well as semantics. This sparkled significant interest in {\em semantic communication}, a novel paradigm that involves the meaning of message into communication. In this article, we first review classic semantic communication frameworks and then summarize key challenges that hinder its popularity. We observe that some semantic communication processes such as semantic detection, knowledge modeling, and coordination, can be resource-consuming and inefficient, especially for communication between a single source and a destination. We therefore propose a novel architecture based on federated edge intelligence for supporting resource-efficient semantic-aware networking. Our architecture allows each user to offload computationally intensive semantic encoding and decoding tasks to edge servers and protect its proprietary model-related information by coordinating via intermediate results. Our simulation result shows that the proposed architecture can reduce resource consumption and significantly improve communication efficiency.
\end{abstract}


\section{Introduction}
\label{Section_Introduction}

Shannon introduced classical information theory in 1949 which first proved reliable communication is possible in noisy channels. In his seminal work\cite{Shannon1948}, Shannon defines the fundamental problem of communication as ``that of reproducing at one point either exactly or approximately a message selected at another point". He argued that the ``semantic aspects of communication should be considered as irrelevant to the engineering problem". The reason is that the meaning of message can be correlated with ``certain physical and conceptual entities" and involving the meaning into the mathematical model may affect the generality of the theory\cite{Shannon1948}. 
Motivated by this principle, most existing communication technologies are developed to maximize data-oriented performance metrics such as 
communication data rate, while ignoring the service/content/semantic-related information or only considering these information in the upper layers (e.g., the application layer). 

In the recent development of wireless technology, the service diversity and service-level optimization solutions based on the content of the message have been embraced by 
both industry and academia. 
More specifically, the latest iteration of wireless technology, the so-called fifth generation (5G), 
has been transformed from the traditional data-oriented architecture to the service-based architecture (SBA) which promises to support a diverse set of services and verticals, some of which can only be enabled by carefully tailored network resources and capacities based on the contents, requirements, as well as semantics of communication. Furthermore, it is commonly believed that 6G will enable more human-centered services and applications such as the Tactile Internet, interactive hologram, and intelligent humanoid robot, which will rely more on the human-related knowledge and experience-based metrics\cite{XY20206GSelfLearn}.  
This raises the question of whether or not the principle of ``semantic is irrelevant" is still necessary for the next generation wireless technology. In particular, a
novel paradigm, referred to as the {\em semantic communication}, which allows the meaning of the message to be sensed and exploited during communication\cite{Weaver1949MathTheoryComm, Juba2010PhDThesisSemanticComm, Bao2011SemanticComm, Guler2018SemanticGame}, has attracted significant interest recently. 
\blu{Compared to the classic communication theory, semantic communication draws inspiration from human language communication focusing on delivering the meaning (e.g., interpretation) of the message which has the potential to fundamentally transform the existing communication architecture towards a more generally  intelligent and human-oriented system.}

Compared to the traditional data-oriented communication frameworks, semantic communication will bring the following unique advantages:

\noindent\blu{\bf Improved Communication Efficiency and Reliability:} It is known that the traditional discrete-channel-based model suffers from low efficiency in some cases. For example, as Shannon argued in \cite{Shannon1948}, ``transmitting a continuous source such as speech or music with exact recovery will require a channel with infinite capacity" and the solution is to discretize the signal within a certain tolerance of information loss, i.e., satisfying certain fidelity requirements. In other words, Shannon theory has a limited efficiency when continuous signal source, especially human-oriented source, is involved. Instead of converting the continuous source signal (e.g., a speech signal) into discrete form with a certain loss of fidelity, semantic communication allows transmission of the meaning of the signal (e.g., the transcript of the speech) which will have the potential to achieve lossless (semantic) information delivery with a significantly reduced demand on communication resource. \blu{For example, delivering a one-hour speech recorded in the form of the voice signal at a rate of 64 kbit/s will require transmitting a voice file with around 230 MB storage. In contrast, transferring the main idea (e.g., summary) of the speech transcript could only need to send a few kbits of text messages. The reliability of communication can also be improved by allowing the decoder to infer the missing part of a corrupted message based on the semantics of the context. Furthermore, it has been shown in \cite{Guler2018SemanticGame} that if the encoding and decoding can be coordinated based on the semantic-related side information such as context and intention of users, the semantic error (difference in the meaning of the source signal and that of the recovered signal) can be significantly reduced.}

\noindent{\bf Enhanced Quality-of-Experience (QoE) for Human-oriented Services:} Traditional communication systems mainly focus on data-oriented metrics including 
data rate and probability of error, none of which reflect the subjective view of the human users. In semantic communication, however, the main objective is to deliver the intended meaning which will depend on 
both the physical content of the message as well as the intention, personality, and other human-oriented factors that could reflect the real QoE of human users. \blu{Consider the hypothetical example in \cite{Bao2011SemanticComm} where a child asks her father what is a ``Tweety". In response to the query, the father will send the answer which may correspond to a yellow canary bird, the client of social media website Twitter, or a character in a cartoon show. By observing the environment that child asks the question as well as the fact that the child may not understand ``canary", the father may choose to send the answer ``Twitter is a bird" to maximize the probability of successful delivery of semantic information.}  

\noindent{\bf Protocol/Syntex-independent Communication:} 
\blu{It is known that contemporary communication systems consist of 
many incompatible communication protocols, e.g., TCP/IP, HTTP, FTP, etc., resulting in continuously growing complexity of the network. Significant effort has been adopted to address the incompatibility-related issues} such as designing protocols that are backwards compatible and introducing new interfaces enabling interoperability as network systems continuing to evolve. Semantic communication built on the common knowledge shared among all the devices as well as human users will lay the foundation for a more robust and upgrade/evolution-friendly and protocol/syntax-independent communication framework for future wireless systems. \blu{For example, suppose signals transmitted with multiple incompatible protocols can be arrived at each receiver. In this case, a receiver does not have to know which protocol associated with each received signal, but can apply different protocols to recover the signal and select the most semantically correct message. It has already been proved in \cite{Juba2010PhDThesisSemanticComm} that by allowing the communication participants to sense the difference between the received signal and the final goal of communication, it is possible to achieve universal communication between any source and destination without requiring a common language/protocol in data communication.} 

Although promising, practical implementation of semantic communication has been hindered by several challenges. In particular, semantic information can be difficult to detect, extract, and represent due to its close relation with background, personality, interaction history, as well as other factors such as the semantic ambiguity (i.e., polysemy) and nuances. Also, detecting and extracting semantic information such as object classification, knowledge entity recognition, and relation inference, often require a significant amount of computing power and storage space, as well as a large number of labeled training data samples, most of which are unavailable in today's network infrastructure. In addition, the semantic information not only consists of explicit information (e.g., color and shape of an object), but also involves unobservable state of the system such as knowledge relations, properties of objects, implicit meaning of statements, etc., which makes it difficult to present and communicate in a simple and elegant form. Also, since the semantic information can be closely related to some highly sensitive human-related information, data and privacy protection will be of critical importance. How to design a simple and general data protection mechanism that can still support collaborative learning and training of a shared knowledge model is still an open problem.

In this article, we first review the classic semantic communication model and then propose a novel architecture based on federated edge intelligence (FEI). In our architecture, users offload resource-consuming semantic processing tasks to the edge servers. Two or more edge servers can also collaborate in training a shared model for processing the common semantic knowledge.  To protect the local semantic data from leakage, we employ a federated learning-based framework in which each edge server cannot expose its local semantic data but can only coordinate with others using intermediate model training results. We conduct extensive simulations to evaluate the performance of our proposed architecture.

\section{Classic Semantic Communication Model}
\label{Section_ClassicModel}

\begin{figure*}
\centering
\includegraphics[width=4.5 in]{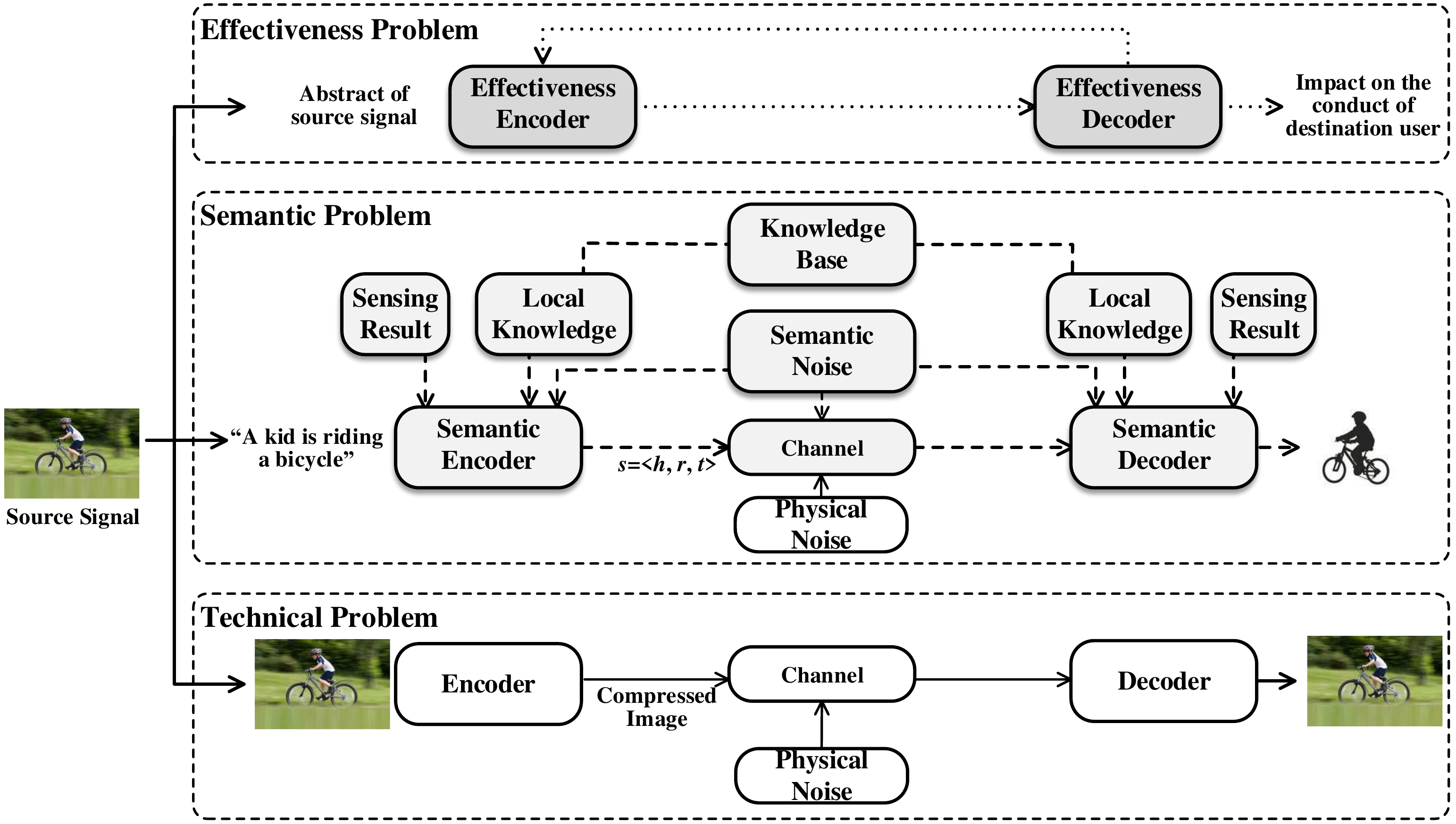}
\caption{Three levels of communication model introduced by Weaver and Shannon.}
\label{Figure_SemCommModel}
\end{figure*}

\subsection{A Basic Semantic Communication Model}
Inspired by Shannon's classic information theory\cite{Shannon1948}, Weaver and Shannon proposed that a more general definition of communication theory should involve three levels of problems listed in sequential order as follows\cite{Weaver1949MathTheoryComm}:
\begin{itemize}
\item[{\bf P1.}] {\bf Technical Problem:} How accurately can the symbols of communication be transmitted?

\item[{\bf P2.}] {\bf Semantic Problem:} How precisely do the transmitted symbols convey the desired meaning?

\item[{\bf P3.}] {\bf Effectiveness Problem:} How effectively does the received meaning affect conduct in the desired way?
\end{itemize}

Classic information theory introduced by Shannon addresses the technical problem by targeting particularly at the accurate transference of source signals to the destination receiver.  
%
The semantic and effectiveness problems must take into consideration much wider research areas including the philosophical content of the communication\cite{Weaver1949MathTheoryComm}, i.e., 
the participating parts in the communication must share the same or similar philosophical worldview such as  
ontology, limits of knowledge, epistemology, logic, and aesthetics. In this article, we mainly focus on the semantic problem of communication.

A semantic communication model needs to recognize and transform the meaning of the source signal into the form that is understandable for both source and destination. For example, suppose the source signal corresponds to an image showing ``a kid named Michael is riding a bicycle", as illustrated in Figure \ref{Figure_SemCommModel}. A technical encoder will ignore the meaning of the image and encode every pixel of the source signal into a message to be recovered by the decoder as accurately as possible. Semantic encoder however will infer the meaning of the source image and coordinate with the destination to make sure the meaning that is desirable for the destination can be delivered. In particular, if the source is able to identify the name of the kid (Michael) which, it believes, is also known by the destination, it will first recognize the entities (e.g., ``Michael" and ``bicycle") as well as their relationship (e.g., ``is riding") and then generate the coded message. On another case that the destination does not know the name of the kid, the source will have to encode message ``a kid named Michael is riding a bicycle" because, in this case, entity ``kid" is the shared knowledge of both source and destination. 

\subsection{Semantic Communication Components}
Weaver argued in \cite{Weaver1949MathTheoryComm} that Shannon's classic information theory is ``general enough to be extended to address the semantic and/or the effectiveness problems of communication". This motivates a series of work\cite{Juba2010PhDThesisSemanticComm, Bao2011SemanticComm, Guler2018SemanticGame} focusing on improving the communication efficiency, e.g., compress the transmit signal and improve the successful rate of the information reception, by introducing additional semantic communication components. 

\noindent{\bf Semantic (Source) Encoder} detects and extracts semantic content (e.g., meaning) of the source signal and compresses or removes the irrelevant information. 
Consider the semantic communication example in Figure \ref{Figure_SemCommModel}. The encoder needs to first identify the entities in the source image based on the local knowledge at the source and destination and then infer possible relationship according to a common world model, e.g., it makes more sense for a kid to ``ride" a bicycle instead of the reverse. 

\noindent{\bf Semantic Decoder} interprets the information sent by the source and recovers the received signal into the form that is understandable by the destination user. The decoder also needs to evaluate the satisfactory of the destination user and decide whether or not the receipt of the semantic information is successful. 


\noindent{\bf Semantic Noise} is the noise introduced during the communication process that causes misunderstanding and incorrect reception of the semantic information. It can be introduced in the encoding, data transportation, and decoding processes. 

\section{Key Challenges for Semantic Communication}
\label{Section_Challenge}


\subsection{Semantic Information Detection and Processing}
%
One of the key prerequisites for semantic communication is to accurately and quickly recognize and extract intended semantic information such as various types of entities and their possible relations and present these information into desired forms.
Unfortunately, the state-of-the-art data clustering, classification, image and voice recognition, and object identification algorithms
heavily rely on large deep learning models that are typically computational intensive and require a large number of high-quality human-labeled data to perform model training. 
Actually, a recent report suggests that the resource consumption of AI solutions is growing in an unprecedented speed and the cost for training an advanced AI algorithm has been doubled every few month and increased over 300,000 times from 2012 to 2017\cite{Schwartz2020GreenAI}. There is still lacking a simple and general solution for quick semantic information detection and processing that can be implemented in resource-limited devices\cite{XY2018EHFogComputing}. 

\subsection{Semantic Knowledge Modeling}
%

Source and destination need to maintain and constantly update its local knowledge models to capture the rich meaning of knowledge entities as well as their complex relationship. One potential solution is to adopt a graphical structure, e.g., a semantic knowledge graph, to model the semantic relationships between different entities. However, a knowledge graph with a large number entities and multi-relational edges can be highly complex and difficult to manipulate. The lack of basic understanding of various semantic structures and meanings of contents further exacerbate the challenge for adopting the semantic knowledge graph in communication systems. 

%

\subsection{Knowledge Coordination and Data Protection}
Depending on the background and interaction history, different devices or users may have different knowledge bases and structures. It is therefore important for different communication participants to coordinate during the communication process. 
However, how to developing a effective coordination mechanism that can enable quick and smooth coordination among communication participants without causing any private data exposure is still an open problem.

\section{A Semantic-aware Networking Architecture}
\label{Section_SemanticNetwork}
In this section, we propose a novel architecture based on federated edge intelligence (FEI) with knowledge/model sharing\cite{ICDCS2021FEI, XY2018TactileInternet, Zhou2019EdgeIntelligence,ChenMZ2021FL} that has the potential to address some challenges mentioned in Section \ref{Section_Challenge}. 

\subsection{Architecture}
Our architecture consists of the following components as shown in Figure \ref{Figure_FEIArchitecture}:
%

\begin{figure}
\centering
\includegraphics[width=3.5 in]{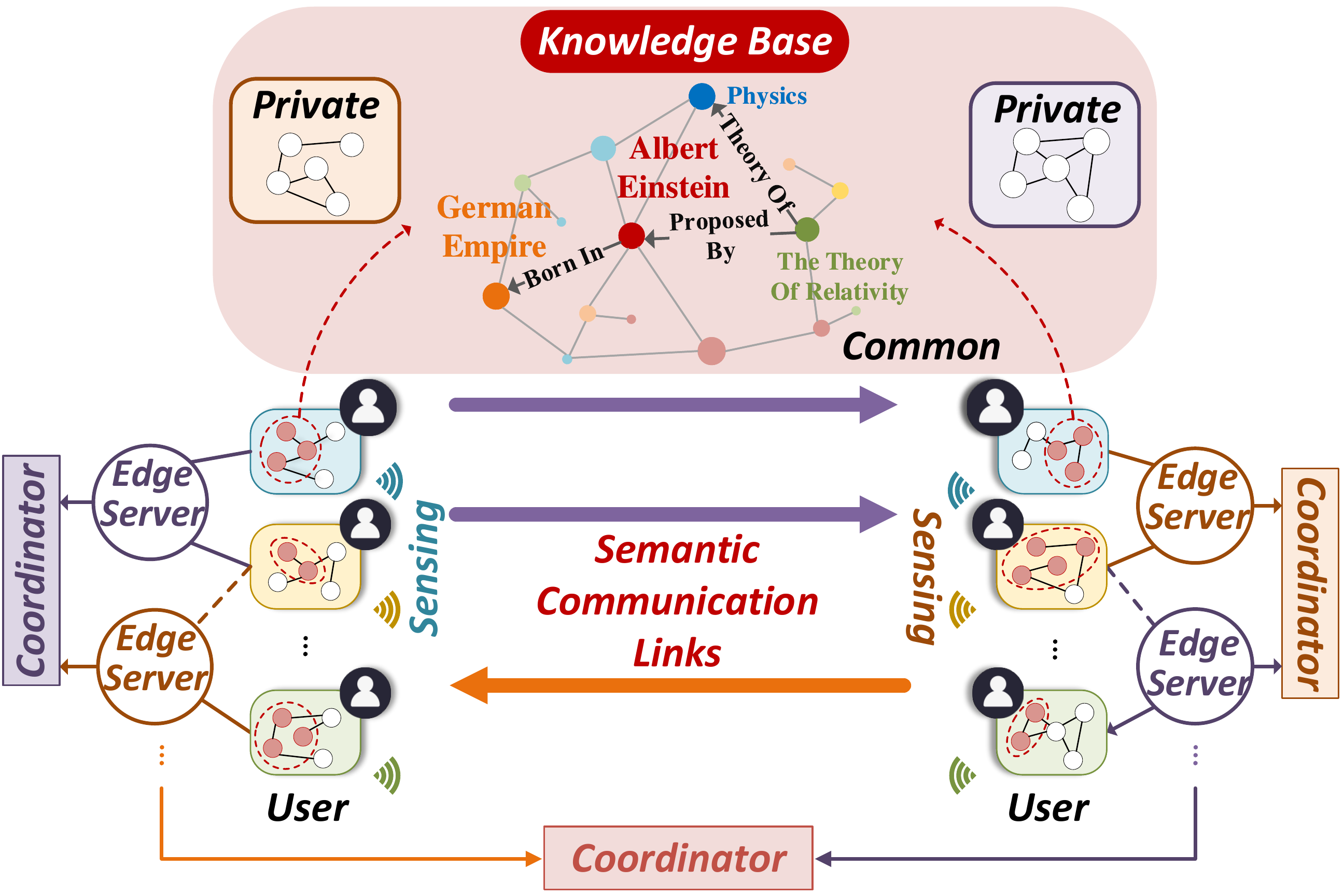}
\caption{\blu{FEI-based architecture for semantic-aware networking.}}
\label{Figure_FEIArchitecture}
\end{figure}

\noindent{\bf Users} correspond to communication participating devices (users) with limited on-board resources. They can either correspond to the low-cost information generators/data collectors ({\em source user}), e.g., IoT devices, sensors, and wearable, or 
be the receivers ({\em destination user}) trying to present the recovered signal consisting of the
intended semantic information to the associated user (e.g., a machine or a human-oriented user). For each source user, it will first sense the specific communication scenario and then upload both the sensing results and the source signal to the closest edge server for knowledge extraction and encoding message generation. For the destination user, it will consult its associated edge server about interpretation of the received signal. 


\noindent\blu{{\bf Knowledge Base} 
consists of elements in real-world knowledge including facts, relations, and possible ways of reasoning that can be understood, recognized, and learned by all the communication participants. Note that the source and destination users do not have to access the same knowledge base. However, the knowledge elements involved in their communication must be known by both users. Consider the example in Section \ref{Section_Introduction}, although the size of the knowledge base of the father is larger than that of the child, the knowledge involved in their communication about the meaning of a Tweety, e.g., a bird, must be involved in the knowledge base of both users. The knowledge base can consist of commonly known facts and relations, e.g., ``Albert Einstein is an expert in physics" as well as private knowledge shared between a specific pair of source and destination users, e.g., ``Michael is riding a bicycle yesterday".
}

\noindent{\bf Edge Servers} perform encoding and decoding based on the commonly shared knowledge base as well as the private signal uploaded from the local users. Generally speaking, each edge server should have already a certain number of well-trained models for object identification and relation inference. \blu{In the FEI-based architecture, multiple edge servers can coordinate to train and update the same machine learning model without exposing their local data samples.}   

\noindent{\bf Coordinators} 
coordinate the model training and learning among collaborative edge servers. They can be deployed at one of the edge server or the cloud data center. In our architecture, different edge servers can establish and maintain shared AI models without exposing any of their local data uploaded from the users. For example, if we adopt FedAvg algorithm introduced in \cite{McMahan2017FLfirstpaper}, each edge server will first train a local model using data collected by its associated users. A selected group of edge servers will then upload the locally trained model parameters to a coordinator for model aggregation. 


\begin{figure*}
\centering
\includegraphics[width=4.5 in]{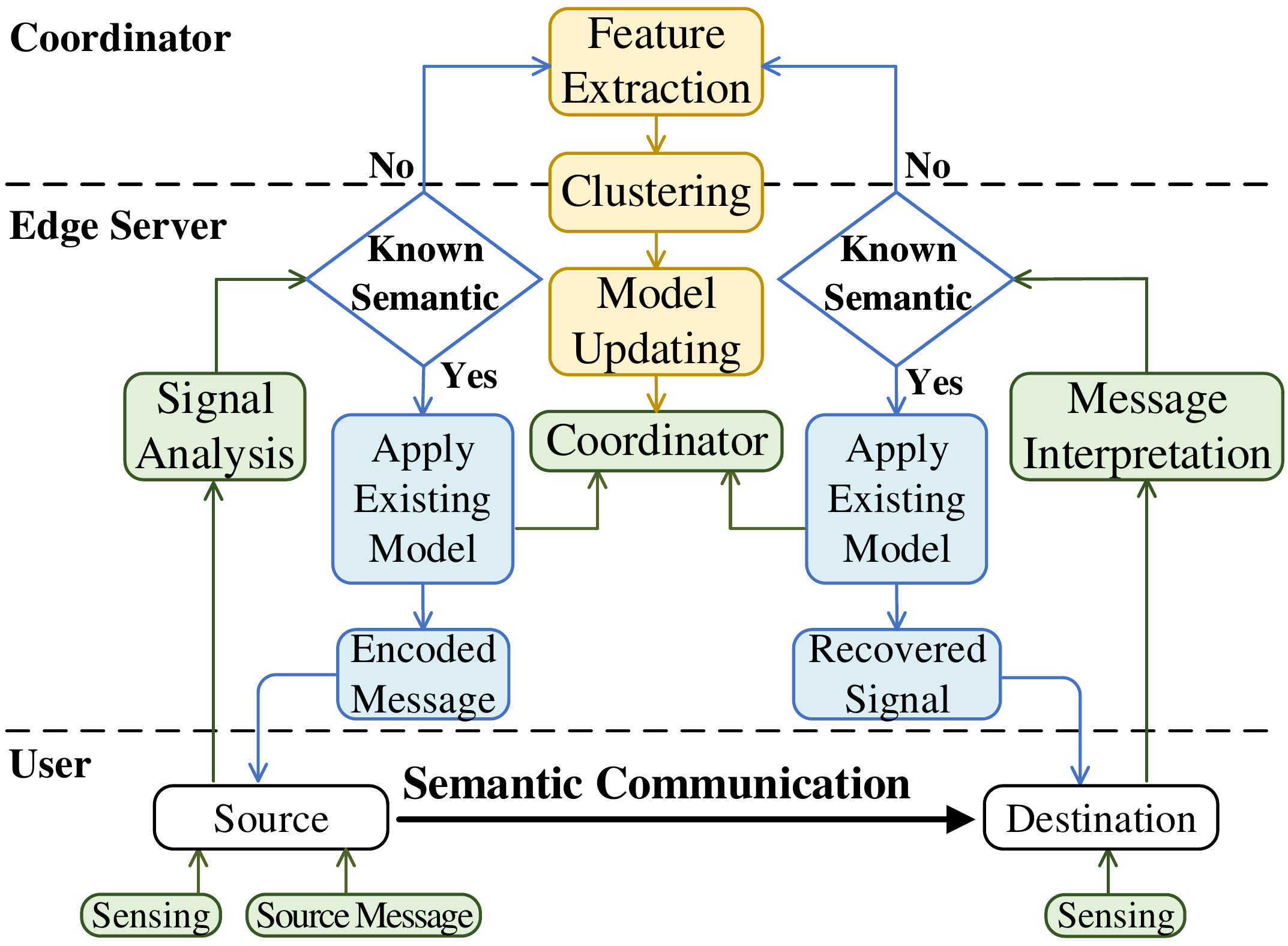}
\caption{\blu{Procedures of the proposed architecture for semantic-aware networking.}}
\label{Figure_SemanticProcedure}
\end{figure*}

\subsection{Key Procedure}
In our architecture, all the above components will interact by performing the following procedures as shown in Figure \ref{Figure_SemanticProcedure}.

\noindent{\bf (1) Sensing:} Each user (source or destination) first senses the surrounding environment to obtain a partial observation signal about the communication environment and scenarios. This signal can be the location (e.g., road, residential, factory, office, etc.) and time stamp (e.g., day or night, peak or idle hour) as well as other information related to the content or semantics of message. 
As will be shown later in this section, allowing both source and destination to sense these information has the potential to significantly reduce the search spaces of knowledge entities and relations (the number of possible knowledge entities and the associated relationship) associated with the content of communication.


\noindent{\bf (2) Communication:} Each source (or destination) user will upload its sensing result together with the source signal (or received signal) to the edge server for semantic encoding (or decoding). The edge server will search for a suitable model to detect the meaning of the signal, e.g., adopt image/voice recognition algorithms to identify the knowledge entities and employ a common/private knowledge graph to estimate the relations among entities. The edge server will also generate the encoded (or recovered) message to be sent by the source user (or destination user for semantic information delivery).

\noindent{\bf (3) Inference:} 
Instead of directly detecting entities and relationship from the signal, encoder and decoder should also have the ability to learn and reason about the possible form of relationship and implications based on the relevant information such as observed environment and scenario, time, locations and social relationship of source and destination as well as their communication history.
%
As will be shown later in this section, due to the high correlation among some semantic knowledge entities, adopting inference at the destination will have the potential to reduce over 90\% of data volume to be transmitted in the networks under certain conditions.

\noindent{\bf (4) Model Updating:} It is possible that some semantic knowledge terms in the signals uploaded from the users cannot be perfectly recognized and processed by the edge servers, e.g., the entities or the relationship identified by the existing models has low likelihood to be correct or reasonable. \blu{In this case, the edge server can first extract features from the unknown elements and then adopt clustering solution (e.g., deep embedded clustering (DEC)) to divide signals into different classes. Each class of signals will be assigned with a new label to be added to the knowledge base and the element recognition model will also be updated.} 

\subsection{Performance Evaluation}

To evaluate the performance of our proposed architecture, we consider the following examples to demonstrate the benefits and costs of semantic communication.



\begin{figure}
\centering
\includegraphics[width=2.5 in]{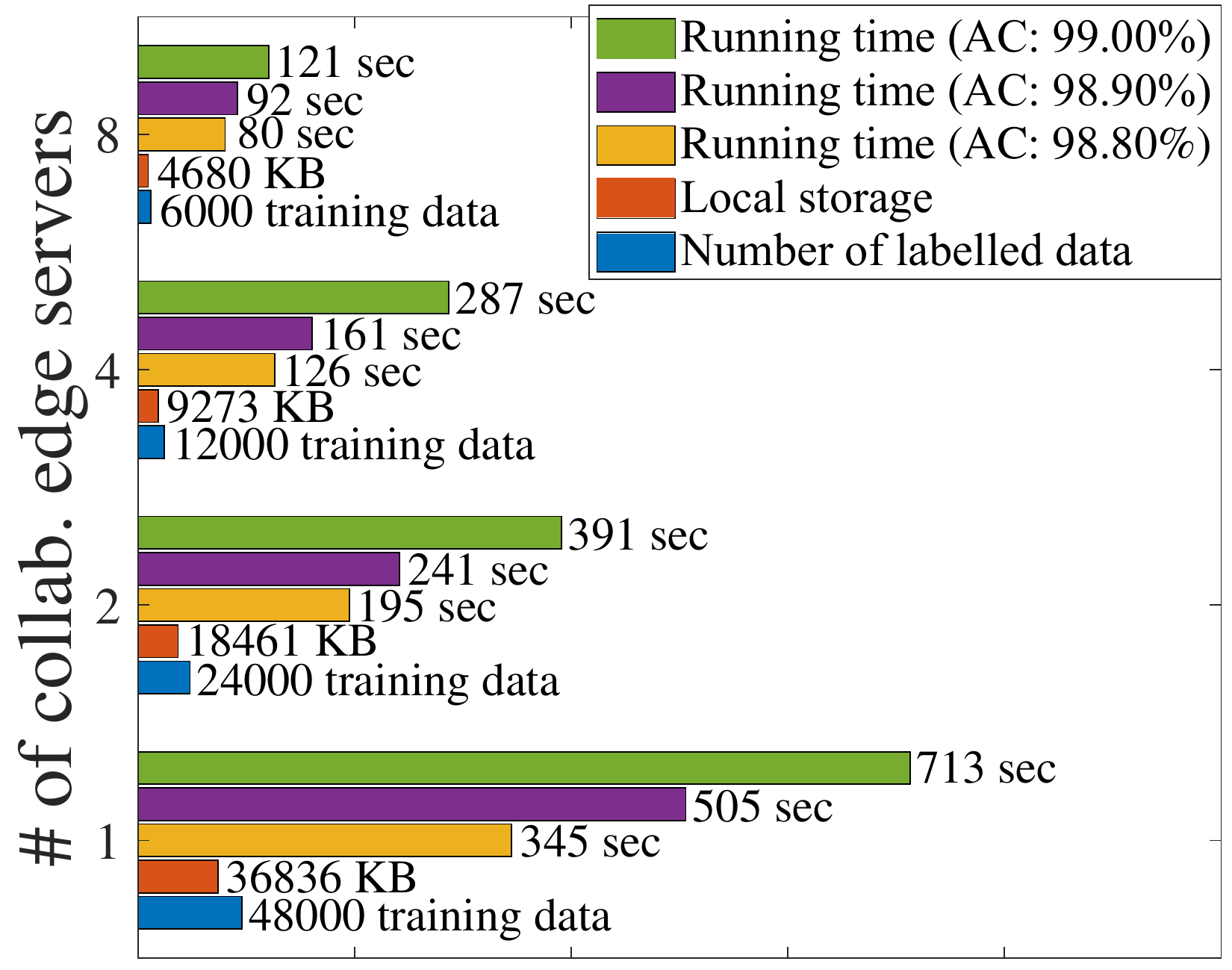}
\caption{\blu{Average running time, local storage sizes, and number of labeled samples per edge server under different numbers of collaborative edge server.}}
\label{Figure_FEI}
\end{figure}

\noindent\blu{{\bf Example 1 -- Source Signal Corresponds to an Image:}
We first consider a simple scenario in which the source signal is in the form of an image file. In this case, the source user can recognize and send the semantic information of the source signal, instead of transmitting the entire image file to the destination user. We use a standard image dataset MNIST as an example. MNIST dataset consists of total 60,000 images of handwritten digits, each has 28 $\times$ 28 pixels corresponding to 6.3 kbit per image. If the semantic information corresponds to the digit in each image which can be encoded with a 8-bit ASCII code, we can observe a significant reduction in transmit file size by using semantic encoding. However, as mentioned earlier in Section \ref{Section_Challenge}, the reduction in communication overhead does not come at no cost. In other words, the source user needs to invest extra infrastructure and resource to perform the image processing. } In Figure \ref{Figure_FEI}, we evaluate the consumption of three types of resources, running time, local storage, and labeled training data samples, for training a simple
CNN model consisting of two $5\times5$ convolutionary layers for recognizing image of handwritten digits based on a standard dataset MNIST. \blu{We use cross entropy as the loss function and the training process terminates when the target accuracy level (98.8\%, 98.9\%, and 99.0\% as shown in Figure \ref{Figure_FEI}) is achieved.}  
Our result shows that, even running on a high-performance GPU server (TITAN X GP102 GPU and Intel i9-9900K CPU@3.6GHz), the time duration for training the AI model would reach up to 345 seconds, 
surpassing the time duration for directly transmitting a high-resolution image file over a wireless channel via any existing wireless technologies including 4G, 5G, and Wi-Fi. We also compare the resource requirement for each individual edge server when two or more edge servers can collaborate in training a shared model with the target accuracy levels\cite{ICDCS2021FEI}. Our result shows that both the required storage and the number of training samples per edge server decrease with the number of model-sharing edge servers and the total running time reduces to less than one fourth of the time with eight collaborative edge servers. \blu{This verifies the effectiveness of our proposed FEI architecture on reducing the resource consumption and achieve the model sharing among edge servers.}  


\begin{figure}
\centering
\includegraphics[width=2.5 in]{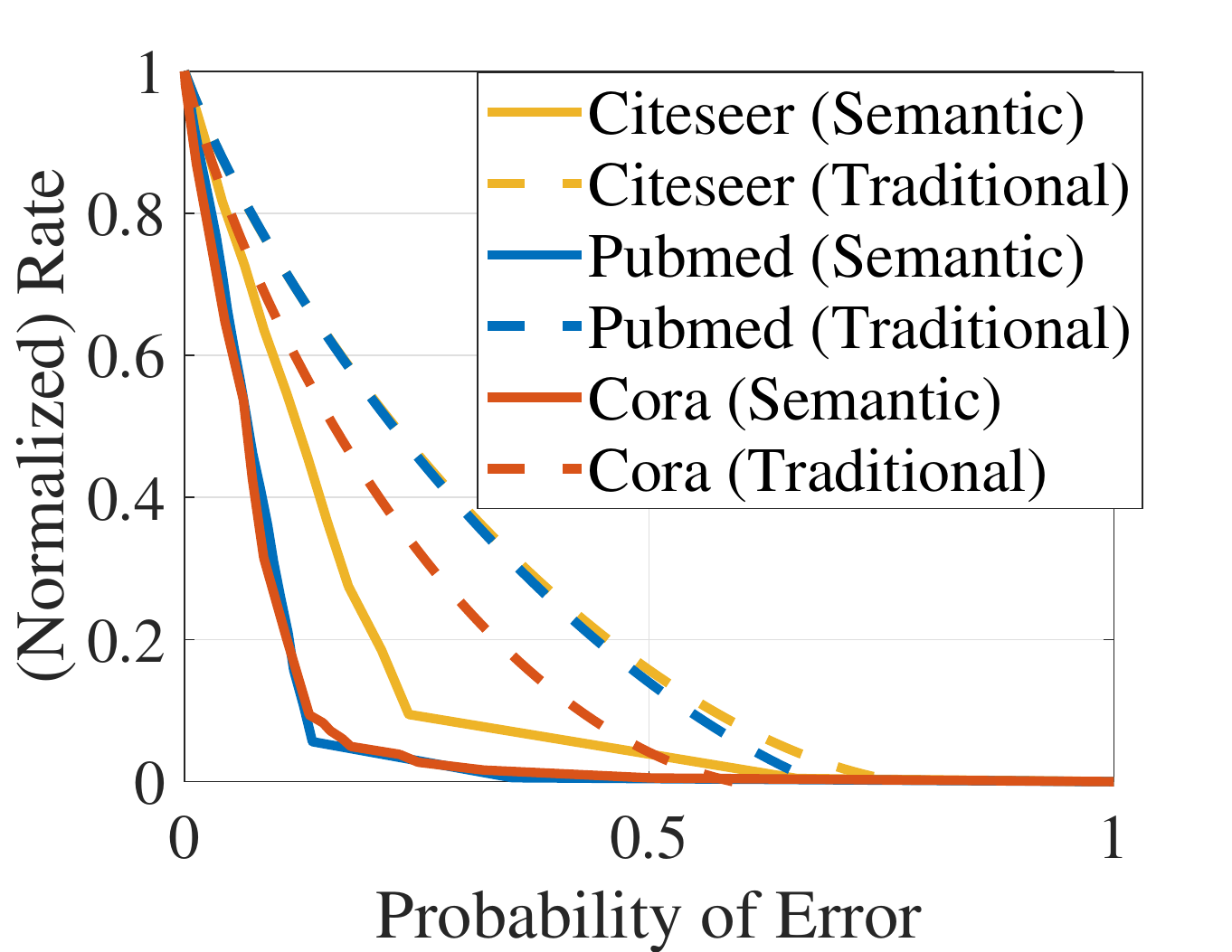}
\caption{\blu{Percentage of labeled data rate input for inferencing entity identification and the corresponding error probabilities of the output.}}
\label{Figure_GCN}
\end{figure}

\noindent\blu{{\bf Example 2 -- Semantic Represented as a Knowledge Graph:}  We consider the scenario that the knowledge base of each user is a large knowledge graph consisting of knowledge entities (e.g., each entity can corresponds to an item/article in an encyclopedia) and relations (e.g., citation relations among different items/articles in an encyclopedia)\cite{Ji2020KnowledgeGraphSurvey} and the semantic information in a source signal can be represented as a subgraph of the knowledge graph.  
We assume the knowledge entities in the subgraph has already been identified by the source user and the main objective is to deliver the correct relations among the entities to the destination. We investigate the distortion rate of the communication channel when a shared knowledge base is available at the source and destination users, compared to the scenario without shared knowledge in which the source user needs to send all the relation information between different entities.}   

\blu{
One of the unique feature of the semantic knowledge graph is that strong correlation may exist among some knowledge entities and the relations. For example, the fact ``Albert Einstein introduced the theory of relativity" is closely related to the fact that ``Albert Einstein is an expert in physics".  This feature can be exploited to further  compress the semantic information of the source signal. Suppose the destination can infer the types of all the knowledge entities of a complete knowledge graph from a limited number of labeled entities using a semi-supervised graph neural network (GCN)-based algorithm\cite{Kipf2017GCN}. In this case, instead of sending all the information of the knowledge subgraph, the source should only transmit a subset of labeled data and the destination can infer the rest of the label information using the semi-supervised GCN approach.
%
%
We compare the compression rate of the GCN-based solution using three citation network datasets: Citeseer, Cora, and Pubmed consisting of documents associated with different areas and citation links between any two documents in Figure \ref{Figure_GCN}. Our results show that the compression rate is closely related to the structural correlation of the dataset. In particular, for Cora and Citeseer, transmitting 18.872 KB and 22.308 KB of labeled data (counting around 24.89\% and 27.94\% of the total labeled dataset corresponding to compression rates of 75.11\% and 72.06\%, respectively) are sufficient to recover the relevant areas of all the documents with  accuracy of recovery at 90\%. In Pubmed, however, to achieve the same accuracy of recovery, it needs to send 135.47 KB of labeled data which is  57.25\% of total labeled dataset resulting in around 42.75\% compression rate. We also present the distortion rate when the destination user does not have any knowledge base but need to receive more relation-based information from the source user to minimize the distance (measured by Hamming distance) between the signal at the source and that recovered by the destination user. We can observe that semantic communication is able to achieve much lower distortion rate in signal recovery at the destination.
}

\section{Open Research Topics}
\label{Section_ChallengesandResearchTopics}
\noindent{\bf Knowledge Evolution Tracking:}
The human knowledge can evolve over time. Modeling and keeping track of temporal variation, e.g., aggregating new knowledge entities and relations and discard obsolete information, of each individual knowledge are helpful to further improve communication efficiency and reduce the probability of error in semantic information delivery. 

\noindent{\bf Network-level QoE Quantification:}  
It is known that people with different ages, genders, personalities, and cultural backgrounds, may have different focuses on their personal experience under different conditions. Developing novel compositional experience metrics covering essential experience indices for diverse types of peoples is an interesting problem worth further investigating.
\noindent\blu{{\bf Capacity of Semantic-aware Network:}
The capacity of semantic networking is more complex and should be closely related to the knowledge sharing among users. 
Developing an elegant  and comprehensive mathematical framework to evaluate  the performance limits of a semantic transportation network is an important direction for future research.}

\section{Conclusion}
\label{Section_Conclusion}
This article proposes a novel architecture based on federated edge intelligence for semantic-aware networking. In this architecture, users can offload the resource-consuming semantic processing tasks to edge servers and 
two or more edge servers can collaborate in training a shared model for processing the common semantic knowledge based on a federated learning-based framework. 
Numerical results show that our proposed architecture can significantly improve the communication performance without causing any local semantic data leakage. Potential topics for future research have also been discussed.
\blu{
The architecture proposed in this paper is far from a complete solution for semantic communication. The main objective however is to identify the potential and challenges to stimulate innovations and developments in semantic-aware networking and applications. We hope our work will spur interests and open new directions on the future evolution of semantic-based networking systems. 
}

\section*{Acknowledgment}
This work was supported in part by the National Natural Science Foundation of China under Grant No. 62071193, 61621005, 61632019, and 61836008, the Key R \& D Program of Hubei Province of China under Grant No. 2020BAA002, China Postdoctoral Science Foundation under Grant No. 2020M672357.

\bibliography{DeepLearningRef}
\bibliographystyle{IEEEtran}

\begin{IEEEbiographynophoto}{Guangming Shi} (SM'06, F'20) received the M.S. degree in computer control, and the Ph.D. degree in electronic information technology from Xidian University, Xi'an, China, in 1988, and 2002, respectively. He is a Professor with the School of Artificial Intelligence, Xidian University. He is also the Vice President of Xidian University. He is an IEEE Fellow and the chair of IEEE CASS Xi'an Chapter, senior member of ACM and CCF, Fellow of Chinese Institute of Electronics, and Fellow of IET. 
His research interest includes Artificial Intelligence, Intelligent Communications, Human-to-Computer/Human-to-Machine Interaction.
\end{IEEEbiographynophoto}

\begin{IEEEbiographynophoto}{Yong Xiao}(S'09-M'13-SM'15) is a professor in the School of Electronic Information and Communications at the Huazhong University of Science and Technology, China. 
His research interests include machine learning, game theory, and their applications in cloud/fog/mobile edge computing, green communication systems, wireless networks, and Internet-of-Things.
\end{IEEEbiographynophoto}

\begin{IEEEbiographynophoto}{Yingyu Li} is a postdoc researcher in the School of Electronic Information and Communications at the Huazhong University of Science and Technology, China. Her research interests include IoT, mobile edge computing, and AI for wireless networks.
\end{IEEEbiographynophoto}

\begin{IEEEbiographynophoto}{Xuemei Xie} is a Professor in the School of Artificial Intelligence at Xidian University.  
Her research interests are human action recognition, object detection, scene understanding, video analysis, deep learning and feature representation.
\end{IEEEbiographynophoto}

%

\end{document}